\documentclass[fleqn,usenatbib]{mnras}

\usepackage{newtxtext,newtxmath}
 
\usepackage[usenames,dvipsnames]{color} %using the color package, not xcolor

\usepackage[T1]{fontenc}

\DeclareRobustCommand{\VAN}[3]{#2}
\let\VANthebibliography\thebibliography
\def\thebibliography{\DeclareRobustCommand{\VAN}[3]{##3}\VANthebibliography}

\usepackage{graphicx}	% Including figure files
\usepackage{amsmath}	% Advanced maths commands
\usepackage{amssymb}	% Extra maths symbols

\newcommand{\ha}{H$\alpha$}
\title[HI Rich Low SF galaxies]{HI Rich but Low Star Formation galaxies in MaNGA: Physical Properties and Comparison to Control Samples}

\author[Sharma et al.]{
Anubhav Sharma,$^{1}$\thanks{E-mail: anubhavprasadsharma@gmail.com (AS), \newline klmasters@haverford.edu (KLM)}
Karen L. Masters,$^{1}$ David V. Stark,$^{1,2}$ James Garland,$^{1}$ Niv Drory,$^{3}$ \newauthor Amelia Fraser-McKelvie,$^{4}$ and Anne-Marie Weijmans.$^{5}$
\\
$^{1}$Departments of Physics and Astronomy, Haverford College, 370 Lancaster Avenue, Haverford, PA 19041, USA\\
$^{2}$Space Telescope Science Institute, 3700 San Martin Drive, Baltimore, MD 21210, USA\\
$^{3}$McDonald Observatory, The University of Texas at Austin, 1 University Station, Austin, TX 78712, USA\\
$^{4}$European Southern Observatory, Karl-Schwarzschild-Straße 2, 85748 Garching, Germany\\
$^{5}$School of Physics and Astronomy, University of St Andrews, North Haugh, St. Andrews KY16 9SS, UK
}

\date{Accepted 4 Sept 2023. Submitted 1 July 2022}

\pubyear{2023}

\begin{document}
\label{firstpage}
\pagerange{\pageref{firstpage}--\pageref{lastpage}}
\maketitle

\begin{abstract}
    Gas rich galaxies are typically star-forming. We make use of HI-MaNGA, a program of HI follow-up for the Mapping Nearby Galaxies at Apache Point Observatory (MaNGA) survey of the Sloan Digital Sky Surveys to construct a sample of unusual neutral hydrogen (HI, 21cm) rich galaxies which have low Star Formation Rates (SFRs); using infra-red color from the Wide-field Infrared Survey Explorer (WISE) as a proxy for specific SFR. 
    Out of a set of 1575 MaNGA galaxies with HI-MaNGA detections, we find {83} ({5\%}) meet our selection criteria to be HI rich with low SFR. We construct two stellar mass-matched control samples: HI rich galaxies with typical SFR {(High SF Control)} and HI poor galaxies with low SFR {(Low HI Control)}. We investigate the properties of each of these samples, comparing physical parameters such as ionization state maps, stellar and ionized gas velocity and dispersion, environment measures, metallicity, and morphology to search for the reasons why these unusual HI rich galaxies are not forming stars. We find evidence for recent external accretion of gas in some galaxies (via high counter-rotating fractions), along with some evidence for AGN feedback (from a high cLIER and/or red geyser fraction), and bar quenching (via an enhanced strong bar fraction). Some galaxies in the sample are consistent with simply having their HI in a high angular momentum, large radius, low density disc. We conclude that no single physical process can explain all HI rich, low SFR galaxies.
 
\end{abstract}


%%%%%%%%%%%%%%%%%%%%%%%%%%%%%%%%%%%%%%%%%%%%%%%%%%

%%%%%%%%%%%%%%%%% BODY OF PAPER %%%%%%%%%%%%%%%%%%

\section{Introduction}

  The question of what quenches star formation (SF) in galaxies is of significant interest to the extragalactic community, as we work to understand the general nature of galaxies. Alongside the accretion of stellar mass via galaxy mergers, galaxies in our Universe grow by the accretion of gas, and by turning that gas into stars. Stars form directly out of molecular hydrogen (H$_2$), but that molecular hydrogen needs to form out of a neutral atomic hydrogen (HI) reservoir. Since HI is an  essential reservoir for star formation, it is not surprising that galaxies with rich HI content tend to have higher star formation rates \citep[SFR; e.g.][]{Doyle2006,Huang2012}, and it is an interesting question to consider why some galaxies with significant gas reservoirs are not actively forming stars. 
   
  While galaxy star formation rates are observed to correlate most closely with the molecular hydrogen gas density, e.g. the classic Schmidt-Kennicutt relationship, \citep{Kennicutt1998}, and on smaller scales in detailed studies including both resolved HI and CO data \citep[e.g. ][]{Schruba2011}, HI is the dominant mass phase of cold gas in galaxies at the current epoch. For example, studies of the COLD GASS sample (which includes both molecular and HI gas content:  \citealt{Saintonge2016}), suggest that the position of galaxies in the global star-formation, stellar mass plane, the so-called “star-forming sequence” or SFS, is driven by the availability of a cold (atomic) gas reservoir. For a very recent review of the importance of cold gas content to galaxy evolution see \citet{SaintongeCatinella2022}. 
    While there is no evidence for a significant population of gas rich quiescent galaxies \citep{SaintongeCatinella2022}, there do exist HI rich galaxies with low SFR. For example some authors have found that nearly all massive quiescent disc galaxies have typical HI masses \citep[e.g.][although \citet{Cortese2020} suggest that this may be due to under-estimates of extended SF in massive HI discs]{Zhang2019}, and even some quiescent early-type galaxies have HI detections \citep[e.g.][]{Grossi2009}. Understanding why these HI rich galaxies are not forming stars at typical levels for their HI content may provide important insight into galaxy evolution in general. A previous analysis of 28 HI-rich low-star-forming galaxies by \citet{Parkash2019} revealed that 75\% of these galaxies had H$\alpha$ emission consistent with being LIERs (low-ionisation emission-line regions which are linked to old stellar populations, or low luminosity active galactic nuclei, AGN, e.g. \citealt{Yan2012}). It was concluded in that work that the presence of HI gas combined with little to no star formation might be a precondition for LIER emission, \citep{Parkash2019}, but they left consideration about the main mechanisms of quenching in the sample for future work.

    Dynamical effects may also suppress star formation in gas-rich systems. \citet{Davis2015} argue that low star formation efficiency (SFE) in a sample of gas-rich early-type galaxies may be due to these objects being in a unique phase of evolution where gas is still in the process of streaming towards their centers. Similar effects on star formation efficiency due to streaming motions are seen in local regions of M51 \citep{Meidt2013}. Furthermore, increased shock/turbulent heating during a merger may keep HI gas warm, suppressing its ability to contribute to star formation \citep{Alatalo2014, Appleton2014}. 
    The presence of a large bulge may also help stabilize gas discs \citep{Martig2009, Saintonge2012}.  
    
    Previous work on quiescent galaxies with high HI content has suggested that the HI is often distributed in a high angular momentum, large radius, low density HI disc \citep{Lemonias2014,Zhang2019}. At these densities the timescales for HI to collapse into H$_2$ and then H$_2$ to stars are prohibitively long, and in some galaxies gas may also take a long time to be transported to the inner, denser parts of discs, depending on the details of internal secular evolution and external tidal nudging. 
    
    In this paper, we investigate a large sample of HI rich galaxies with low SFR in order to attempt to identify the physical reasons why the high HI content is not able to sustain active star formation. We investigate the dominant ionizing source of the hydrogen in the galaxies using resolved optical spectroscopy from the MaNGA survey. We also investigate other measures revealing the physical condition of these unusual galaxies, which we compare with two sets of more typical galaxy control samples ({HI rich and star forming galaxies and HI poor and low star forming galaxies}). 
    
    Where physical units are employed we assume a flat $\Lambda$CDM cosmology with $H_0=70 {\rm km~s}^{-1} {\rm Mpc} ^{-1}$.\footnote{This is for MaNGA DAP and HI-MaNGA values. Pipe3D quantities are calculated with $H_0=71 {\rm km~s}^{-1} {\rm Mpc} ^{-1}$, however this introduces at most a 3\% systematic, well within our typical statistical error, and we always do relative comparisons between our sample and mass matched controls.}

\section{Methods}

\subsection{MaNGA Data and Data Products}
    The sample we use in this work is a subset of galaxies observed by the Mapping Nearby Galaxies at Apache Point Observatory (MaNGA) sample \citep{Bundy2015}. MaNGA is a program under the fourth phase of the Sloan Digital Sky Surveys (SDSS-IV; \citealt{Blanton2017}) which has mapped the composition and kinematics of a sample of 10,010 nearby galaxies by use of a IFU (Integral Field Unit) on the Sloan Foundation 2.5m telescope at Apache Point Observatory \citep[for more details on MaNGA instrumentation, telescope and survey strategy see: ][]{Gunn2006, Smee2013, Drory2015, Law2015, Law2016, Yan2016a, Yan2016b}. MaNGA IFU bundles range from 12\arcsec--32\arcsec ~in size (with about 30\% of the complement being the largest size). The MaNGA sample selection is described in \citet{Wake2017}, which explains the primary and secondary samples (designed to have bundle coverage to 1.5$r_e$ and 2.5$r_e$ respectively, where $r_e$ is the effective radius of the galaxy); we use data from the internal release labeled MPL-11 (MaNGA Product Launch-11), which is identical to the final sample that was made public in the SDSS-IV DR17 release \citep{DR17}. 

   MaNGA provides fully reduced data IFU data cubes (via the Data Reduction Pipeline, or DRP; \citealt{Law2016}), as well as some higher order analysis (maps of velocities, emission lines etc) via the Data Analysis Pipeline (or DAP; \citealt{Westfall2019}), however further analysis is needed to obtain star formation rates, stellar masses and other parameters of the stellar population. In this work we make use of the DR17 version {\tt v\_3\_1\_1} of the Pipe3D analysis of MaNGA data which provides such estimates, along with measurements of the ionized gas (e.g. metallicities). The Pipe3D method as applied to MaNGA data is described in \citet{Sanchez2018}. Specifically from Pipe3D we make use of the integrated stellar mass (integrated in the bundle). While a global value of stellar mass might be a better choice to ensure consistency across samples, we find that there is a consistent difference of 0.24dex between the Pipe3D and a global measure of stellar mass from the NASA Sloan Atlas (NSA; \citealt{Blanton2011}), a value which is largely explained by differences in Initial mass function (IMF) choice, as explored in detail in \citet{Stark}. In addition, the distribution of bundle sizes, measured in terms of galaxy effective radius, $r_e$, is statistically indistinguishable for the three samples introduced in Section \ref{sec:sample}, so we choose to use these stellar masses from MaNGA data directly in our comparison.
   
   From the MaNGA DAP we make use of
   \begin{itemize}
       \item \ha ~``rotation speed", asymmetry and dispersion from the Gaussian emission line fit in the DAP. We define the rotation speed measure to be the separation between the {\tt HI\_CLIP} and {\tt LO\_CLIP} (from the DAP these are the 97.5\% and 2.5\% percentile values in the MaNGA bundle respectively, found after clipping $3\sigma$ outliers), while the velocity asymmetry is the difference in their absolute values. For velocity dispersion we use the average value of dispersion measured by the DAP within 1$r_e$. This is not corrected for instrumental effects (see \citealt{Law2021} for a discussion of the need to do this correction), but we are only using it to compare with control samples, so all should have similar corrections. 
       \item Stellar velocity ``rotation speed", asymmetry and dispersion are taken from the DAP stellar velocity map using similar definitions as for the \ha ~ velocity measures above.
   \end{itemize}
   Note that we say ``rotation speed" as we have not visually checked that these galaxies show symmetric rotation - so this is the equivalent of half the width of a histogram of velocity measures in these maps. 
   
   MaNGA targets are selected from the NASA Sloan Atlas (NSA; \citealt{Blanton2011}), and various NSA values are provided in MaNGA tables. We make use of $b/a$, the isophotal axial ratio from elliptical Petrosian analysis from the NSA as tabulated in the MaNGA DAP summary table.  
    
    Average gas-phase metallicities, $12+\log{O/H}$, within 1 $r_e$ are estimated using the N2O2 strong-line method of \citet{Kewly2002}, which uses the  [\ion{N}{ii}]6585{\AA}/[\ion{O}{ii}]3727{\AA} flux ratio. Internal extinction corrections are performed on each spaxel using the Balmer decrement, requiring both the H$\alpha$ and H$\beta$ lines to have flux $S/N>3$. 12+log(O/H) is calculated in each spaxel, then averaged within 1 $r_e$. We assume a Balmer decrement of 2.86 \citep{Osterbrock2006} and an extinction curve of \citet{ODonnell1994}. Although there are significant systematic errors in the metallicity zero point of different strong-line methods, relative comparisons of metallicity, like those performed in this study, should be largely robust against the indicator used \citep{Kewley2008}.

   \subsubsection{Ionization Analysis}
There are clear trends between the global properties of galaxies and the dominant ionizing source \citep{Sanchez2020}. Generally, the high energy photons for the ionisation of hydrogen comes either from the young stars generated in active star formation or from infall of material onto an active galactic nuclei (AGN); however other ionization mechasims exist (e.g. the exposed cores of old stars, or shocks). These different mechanisms create different line ratios of common species. The \citet[][BPT]{Baldwin1981}  optical line ratio diagnostic allows astronomers to differentiate whether the ionisation comes from star formation or AGN, or old stellar populations, or shocks (these latter two generally combined together as "Low Ionization Emission Regions", or LIERs).

\begin{figure*} 
    	\includegraphics[width=\columnwidth]{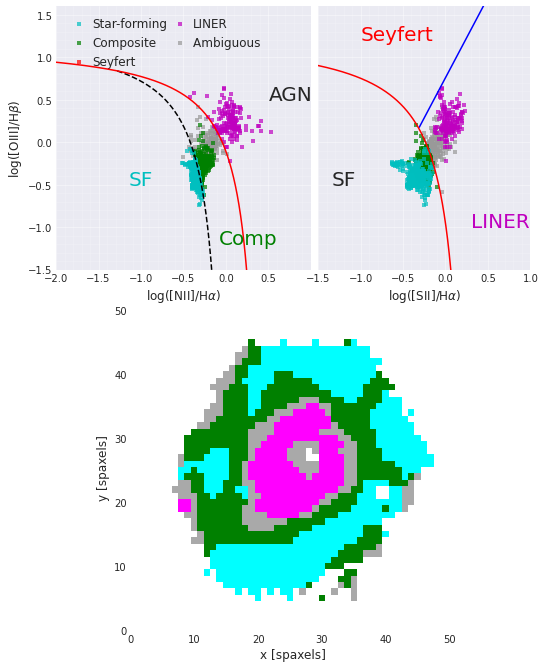}
    	\includegraphics[width=\columnwidth]{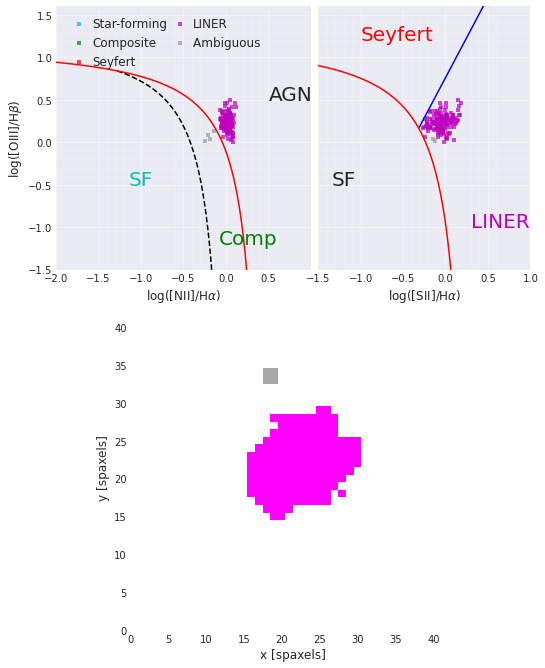}
    	\includegraphics[width=\columnwidth]{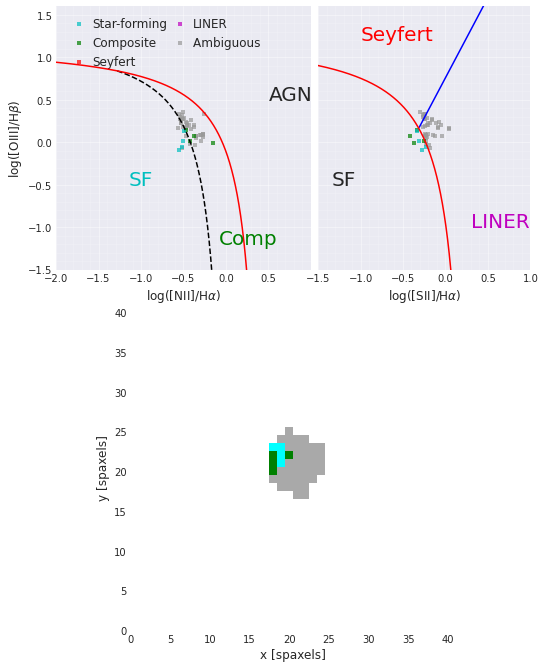}
            \includegraphics[width=\columnwidth]
        {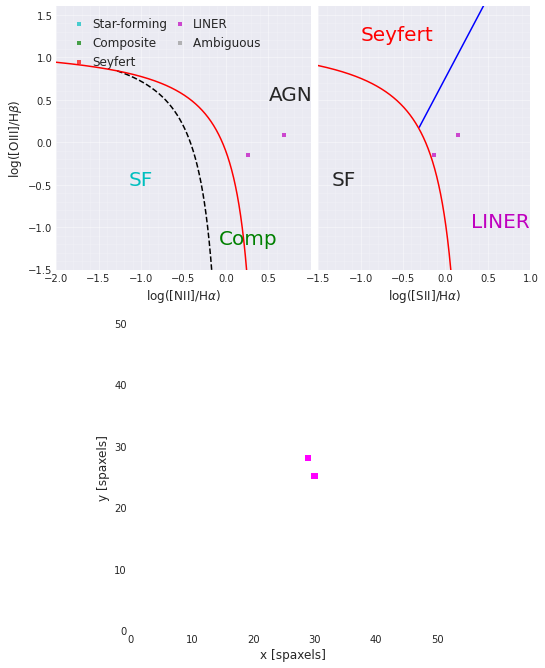}
        \caption{Example BPT diagrams {(generated using the {\it Marvin} tool {\tt get\_bpt})} of four different galaxies in our {HI Rich Low SF} sample, illustrating some of our BPT classifications: In cLIER galaxies (example to the top left; {plateifu: 8450-6101}), the central LIER emission is spatially extended, but accompanied by star formation at larger galactocentric distances, while in eLIER galaxies (example to the top right; {plateifu: 9090-3702}), LIER emission is extended throughout the whole galaxy. We find that our sample of HI Rich Low SF galaxies show a greater proportion of cLIERs than are found in {HI Low Control (HI poor galaxies with low SF)}. Galaxies dominated by Ambiguous spaxels (i.e. spaxels who show emission line ratios with ambiguous BPT classification) were classified as {``Composite+"} (example to the bottom left; {plateifu: 9892-3702}) and galaxies with either few emission line spaxels, or spaxels with no clear spatial distributions of classifications were classified as ``Can't Tell" (example to the bottom right; {plateifu: 9088-6103)}}
        \label{fig:bpt}
    \end{figure*}

To generate BPT diagrams for the sample, we make use of {\it Marvin} \citep{Cherinka2019}, a python package designed to work with MaNGA data. Fig.~\ref{fig:bpt} shows two examples of BPT diagrams generated using {the {\it Marvin} tool {\tt get\_bpt}\footnote{\url{https://sdss-marvin.readthedocs.io/en/latest/tools/bpt.html}}}. {\it Marvin} uses demarcation lines and classification from \citet{Kewley2006} to classify each spaxel with enough line emission (we use the default $S/N>3$ in all lines in Marvin, but other choices are possible) as either star-forming (cyan), LINER (pink), Seyfert (red), composite (green) or ambiguous (grey). We visually inspect the BPT plots and manually categorized each galaxy using the morphology of the different spaxel classifications. We based our classifications on those described in \citet{Belfiore2016}, in summary:  
\begin{itemize}
    \item Starforming: if ``most" spaxels are starforming 
    \item cLIERS: showing central Low-ionisation emission-line regions surrounded by other spaxel types (see left side of Figure \ref{fig:bpt})
    \item eLIERS: Extended Low-ionisation emission-line regions (see right side of Figure \ref{fig:bpt})
    \item Seyfert: dominated by Seyfert spaxels
    \item {Composite+ : dominated by Composite or Ambiguous spaxels (i.e. ``most" spaxels ionized by mix of SF and other ionization)}
    \item Can't tell: anything else
\end{itemize}

   \subsection{HI-MaNGA}
   The source of our HI measurements in this work is the HI-MaNGA survey \citep{Masters2019,Stark}, which is making use of the Robert C. Byrd Green Bank Telescope (GBT) to obtain global HI 21cm line measurements for all MaNGA galaxies $z<0.05$ to a comparable depth as the ALFALFA (Arecibo Legacy Fast Arecibo L-band Feed Array) survey \citep{Haynes2018}, and using ALFALFA data in the part of the sky where MaNGA and ALFALFA overlap. We use the DR3 sample from HI-MaNGA \citep{Stark}, which contains information about HI observations of 6632 unique MaNGA galaxies. These data provide the total HI content and HI line width for MaNGA galaxies, however as the beams of the radio telescopes at 21cm are significantly larger than the size of a typical MaNGA galaxy (3\arcmin for Arecibo, and 9\arcmin for GBT compared to $<1\arcmin$ for most MaNGA galaxies) sometimes HI signals from neighbouring galaxies can be confused. For example, C. Witherspoon et al. (in prep.) find that some low mass AGN which have catalogued HI detections are actually HI poor with nearby HI rich companions. \citet{Stark} provide an estimate of the likelihood of that considering both distance to and redshift and colours of nearby galaxies. A test of this confusion algorithm was provided by \citet{Shapiro2022} who conclude that it is a conservative cut - i.e. it removes all potentially confused targets (and some which are not confused). To be sure that our sample of HI rich, low SF galaxies genuinely have their own HI, we therefore use this process to created a clean ``unconfused" sample.
 
   \subsection{WISE, Galaxy Zoo and Galaxy Environments}\label{sec:other}
   
    We also make use of data from the Wide-field Infrared Survey Explorer (WISE), which mapped the sky at 3.4, 4.6, 12, and 22 \(\mu\)m (W1, W2, W3, W4) with an angular resolution of 6.1", 6.4", 6.5", \& 12.0" in the four bands \citep{Wright2010}. These  data can be used to measure star formation of our galaxies in a way which is more robust to the effect of dust obscuration than optical or UV measures of star-formation although there are some disadvantages, such as it missing SF in the outer regions of some galaxies, and also cases where the emission from evolved stars can dominate W3 masking the signal of low levels of SF. However, galaxies with a colour W2-W3 < 2.0 have been demonstrated to have a specific SFR (sSFR = SFR/M$_\star$) $< {10}^{-10.4} {\rm yr}^{-1}$ \citep{Parkash2019}. We make use of magnitudes published in the AllWISE Source Catalog\footnote{https://wise2.ipac.caltech.edu/docs/release/allwise/}, choosing the profile-fit photometry galaxies which are small relative to the WISE point spread function (PSF; i.e. point-sources), and extended aperture photometry for larger sources. This procedure is recommended by the WISE documentation which states that profile-fit photometry will underestimate fluxes for extended sources, which include 79\% of the MaNGA sample. 
    
    We make use of visual morphologies from the Galaxy Zoo (GZ) analysis of SDSS legacy images \citep{Willet2013}. Galaxy Zoo provides quantitative measures describing various different morphologies, allowing us to test both how the bulk morphology (spiral galaxy or elliptical galaxy) and internal features (like spiral arms, or bars) impact the properties of the sample. We use the weighted vote fractions from Galaxy Zoo with redshift debiasing described in \citealt{Hart2016}. These numbers are referred to as morphological feature likelihoods, denoted $p_{\rm feature}$ in what follows.
    
     Finally, we make use of the Galaxy Environment for MaNGA (GEMA) Value Added Catalog \citep[Version 2.0.2][]{Argudo-Fernandez2015, Etherington2015, Wang2016} to investigate both the local density and overdensities of our samples. Out of GEMA, we make use of 
     \begin{itemize}
         \item Local\_density (from HDU14):  the average density of galaxies in a volume out to the 5th nearest neighbour, corrected by mean over-density using the correction in \citep{Etherington2015}.
         \item $Q_{\rm LSS}$ (in HDU5) which describes the relative tidal forces caused by large scale structure within 1 Mpc \citep[see][]{Argudo-Fernandez2015}. 
     \end{itemize}
     
     In addition we calculate a $Q$ parameter (based on that described in \citealt{Verley2007}) that attempts to quantify the tidal force experienced by a galaxy from it's nearest neighbours, within $\Delta V = 500 {\rm km~s}^{-1}$  (J. Garland et al. in prep).
    
\subsection{Sample and Control Sample Selection}\label{sec:sample}
     We cross match the DR17 (v2.7.1) Pipe3D catalog \citep{Sanchez2018}; the HI-MaNGA DR3 catalog \citep{Stark} and data from the Wide-field Infrared Survey Explorer (WISE) catalog for the DR17 sample. We remove any MaNGA bundles which aren't galaxies, or are pairs of galaxies, or have DAPqual flag set to ‘CRITICAL’, as well as any HI detections which fail the test for possible confusion. This results in a parent sample of 4316 galaxies, out of which, 1575 had HI detections and 2741 have HI upper limits. 

\begin{figure*}
    	\includegraphics[width=\columnwidth]{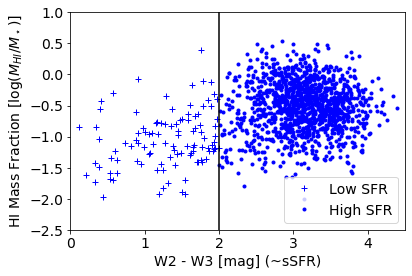}
    	\includegraphics[width=\columnwidth]{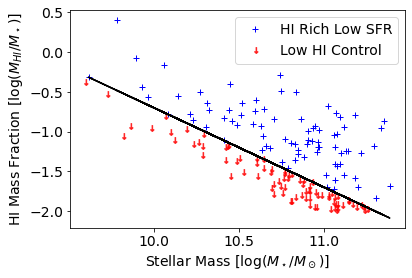}
    	\caption{Left: HI mass fraction plotted against W2-W3 from WISE (a colour which correlates well with sSFR). Our {prospective} sample galaxies are marked with plus-sign, while all other HI detections are shown round points. Right: HI Mass fraction vs stellar mass plot for the main sample and the Low HI control sample (which are shown plotted at the estimated upper limit of undetectable HI). The diagonal line shows the largest upper limit ($\log{M_{HI}/M_\odot}<9.3$) we allow in the low HI control. \label{fig:scatter}}
    \end{figure*}

\begin{figure}
        \includegraphics[width=\columnwidth]{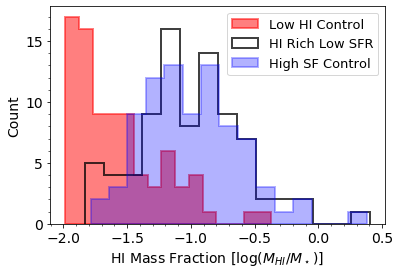}
    	\includegraphics[width=\columnwidth]{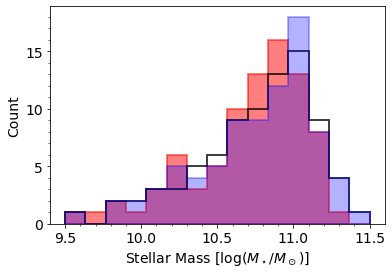}
    	\includegraphics[width=\columnwidth]{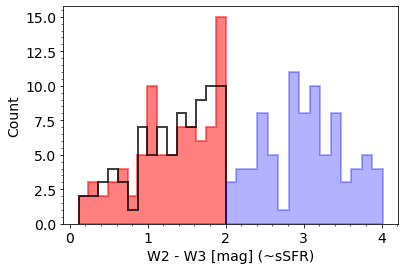}
        \caption{Top: HI Mass Fraction distribution (we show limits for Low HI Control). The main sample and HI rich control have similar HI mass fractions, while the HI control is offset. Center: the distribution of stellar mass for the main sample and both controls demonstrating how they overlap well.  Bottom: The distribution of W2-W3 (as a proxy for sSFR) for all three samples showing the split between high and low SF. See Table \ref{tab:ks_table} for p-values comparing these distributions.}
        \label{fig:sample}
    \end{figure}

     We define all of these 1575 galaxies detected by HI-MaNGA as ``HI rich"; detection in HI-MaNGA typically means they have an HI mass,  $M_{\rm HI} > 10^9 {\rm M}_\odot$ \citep{Masters2019,Stark}. Note that this sample has stellar masses which range from $10^9-10^{11} {\rm M}_\odot$ (see Figures \ref{fig:scatter} and \ref{fig:sample}).  

     To define high and low SFR subsets we make use of the WISE photometry. We follow \citet{Parkash2019} in assuming that galaxies with a colour W2-W3 < 2.0 have a specific SFR (sSFR = SFR/M$_\star$) $< {10}^{-10.4} {\rm yr}^{-1}$. \citet{Parkash2019} based this on the W1-W2 stellar mass calibration of \citet{Cluver2014} with the W3-SFR calibration of \citealt{Brown2017}). We define this to be the limit for the low sSFR sample. This method is more robust to the effect of dust obscuration than optical or UV measures of star-formation although there are some disadvantages to be recalled in interpreting our samples, such as it missing SF in the outer regions of some galaxies, and also cases where the emission from evolved stars can dominate W3 masking the signal of low levels of SF.
     The left panel of Figure \ref{fig:scatter} shows the W2-W3 versus the HI Mass Fraction for these 1575 galaxies with HI detection. We find that 105 ($7\pm 1$ \%) have W2-W3$<2.0$. {Out of these 105 galaxies, we select the 83 galaxies with $\log{M_{HI}/M_\odot}>9.3$ (see right panel of Figure \ref{fig:scatter} for reference) as} our ``Main Sample" of HI rich galaxies with Low SF. 
     {We apply this selection of $\log{M_{HI}/M_\odot}>9.3$ in order to ensure that our HI Rich Low SF sample does not share parameter-space with HI non-detections in the ``Low HI Control".}
     We note that higher mass galaxies typically have lower sSFR even when starforming, so a constant cut in sSFR selects low SF massive galaxies which are somewhat closer to the star-forming population than among the lower mass galaxies.

    
    In order to understand in which ways the low SFR HI rich galaxies differ from more typical galaxies, we create two sets of control samples. Both of them  include galaxies which are selected to have a distribution of stellar masses matched with the stellar mass of the main sample. A ``High SF Control" is selected to have similar HI Mass Fraction and Stellar Mass as our main sample, but W2-W3>2.0 (or sSFR$>10^{-10.4}{\rm yr}^{-1}$). A total of 1330 galaxies meet these selections, out of which we find the {83} with closest matches in stellar mass and HI mass fraction to our main sample, without repeating any galaxies. A ``Low HI Control" is selected to have both low SF (W2-W3<2.0) and similar stellar masses as our main sample, but be HI poor. We use measured upper limits to define this HI-poor comparison set using a threshold of $M_{\rm HI} < 10^{9.3} {\rm M}_\odot$ which picks out galaxies with low HI mass fractions for their stellar mass (see right panel of Figure \ref{fig:scatter}). There is some overlap in the HI upper mass limits and some of our lower mass HI detections, however overall this selection results in a sample with significantly less HI than our main sample. We find 1400 galaxies meeting these selections, and follow the procedure noted above to find the best possible stellar mass and sSFR (color) matched sample. 
    
    {While} the statistical rigour of our work could be improved by implementing multiple random draws for these control samples, {it is important to realize that} even though MaNGA is the largest currently available IFU sample with 10,010 galaxies in total, at this time of analysis, the available HI-MaNGA sample contained $N=6632$ of the full MaNGA sample  {which limits} the number of possible independent controls  {to} less than ten, even if we {allow matches as large as} 0.4 dex within the two matched quantities for each control.
     
    Inclination can impact measures of sSFR via dust obscuration (although by using WISE colours we limit this impact). For disc galaxies, axial ratio is a proxy for inclination. We note that we have applied no morphology selection yet, however a check of the distribution of axial ratios for our three samples, finding statistically similar distributions peaking at $b/a=0.7$, with a hint that the {Low} HI, low SF skews rounded (see Section \ref{Sec:morphology} where we discuss how this subset are, perhaps unsurprisingly, less likely to be disc galaxies so it is reasonable to expect they appear on average rounder on the sky).

    For a comparison of all selection properties of the main sample with both controls please see Fig.~\ref{fig:sample} which shows the histograms of W2-W3, Stellar mass and HI mass fraction as well as plots of HI mass fraction against both W2-W3 and stellar mass. The typical properties of galaxies in the three samples are also summarized in Table \ref{tab:samples}. 
    
    \begin{table}
	\centering
	\caption{Summary of Main Sample and Both Control Samples. All have N={83} galaxies. sSFR limits are based on W2-W3 colours}
	\label{tab:samples}
	\begin{tabular}{lccc}
		\hline
		Sample & $M_{\rm HI}$  & sSFR  &  $\langle \log M_\star/M_\odot \rangle$ \\
		& [${\rm M}_\odot$] & [${\rm yr}^{-1}$] &  \\
		\hline
		Main Sample  &  {$> 10^{9.3}$}  &    $< {10}^{-10.4}$  & {10.76} \\
		(HI Rich, low sSFR) \\
		High SF Control &  {$> 10^{9.3}$} &  $>{10}^{-10.4} $ & {10.76} \\
		(HI Rich, high sSFR)\\
		Low HI Control &  $< 10^{9.3}$ &  $<{10}^{-10.4} $ & {10.72}\\
		(HI Poor, low sSFR)\\
		\hline
	\end{tabular}
    \end{table} 

\section{Results}

\subsection{Source of Ionization}\label{sec:ionization}
    
    In their analysis of 91 HI galaxies with little or no star formation from Parkash (28 of which had Wide Field Spectrograph (WiFES) IFU data), \citet{Parkash2019} discovered that they had very high LIER fractions and speculated that the presence of HI gas with little/no SF may be a precondition for LIER emission. Looking at the resolved BPT diagrams of the samples from MaNGA, we find that our HI {Rich} Low SF (Main) and Low HI Control samples both have a high proportion of LIERs ({$77\pm 11$\%}  and {$60\pm 10$ \%}  of classifiable maps respectively). In contrast, {the High SF Control} was found to have relatively low proportion of LIERs ({$19\pm 5$\%} of classifiable maps).  We thus conclude that in our samples, the high LIER fraction correlates with low SFR, not requiring a combination of low SFR and HI content.

    \begin{figure}
      	\includegraphics[width=\columnwidth]{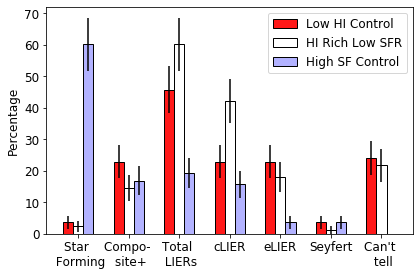}
        \caption{Bar Graph showing the primary ionisation map category of each of the three samples expressed as a percentage. It is clear that both low SFR samples, regardless of HI status, have high LIER fractions.}
        \label{fig:ionization-results}
    \end{figure}
    
    We further categorize LIER emission based on its spatial distribution, following the definitions of \citet[][see Section \ref{sec:ionization} where we describe this in more detail]{Belfiore2016}. Fig.~\ref{fig:ionization-results} shows a visualisation of the primary ionisation category based on the resolved BPT diagrams; numbers are given in Table \ref{tab:bpt}. In cLIER galaxies (for an example see the upper left panel of Figure \ref{fig:bpt}), the central LIER emission is compact, but typically larger than the PSF and is accompanied by star formation at larger galactocentric distances, while in eLIER galaxies (example in the upper right panel of Figure \ref{fig:bpt}), LIER emission is extended throughout the whole galaxy. {Among the LIERs, while we found an equal proportion of cLIERs ($50\pm11.5\%$) and eLIERs ($50\pm11.5\%$) in the Low HI Control, LIERs in both the main sample and High SF Control are skewed heavily towards cLIERs ($70 \pm 12 \%$ and $81 \pm 23 \%$ respectively) compared that to eLIERs ($30 \pm 8\%$ and $19 \pm 11\%$ respectively). This implies that while overall the presence of LIER emission is tied to low SFR (regardless of HI content), cLIER emission also favours high HI content (regardless of global SF).} This excess of cLIERs could be a feature of our W2-W3 based selection for low SF galaxies - which if dominated by a central old stellar population, may still be surrounded by an extended SF disc. Further analysis of the UV of these samples would be interesting.

  \begin{table}
	\centering
	\caption{Summary of Resolved BPT Classifications. Columns show total numbers, and percentages relative to the number of classifiable maps, with Poisson counting errors. Where there is a $^*$ following the percentage we show the percent relative to all LIERs rather than all maps. {Where there is a $^\dagger$ following the percentage we show the percent relative to sample size rather than classifiable maps.}}
	\label{tab:bpt}
	\begin{tabular}{lccc}
		\hline
		Classification & {HI Rich Low SF}  & {High SF}  & {Low HI}  \\
		 & {Sample} & {Control}  & {Control} \\
		\hline
	SF & { 2 ($3\pm2\%$)} & {50 ($60\pm9\%$)} &  {3 ($5\pm3\%$)} \\
	{Composite+} & {12 (19$\pm$5\%)} & {14 (17$\pm$5\%)} & {19 (30$\pm$7\%)} \\
	Total LIERs & {50 (77$\pm$11\%)} & {16 (19$\pm$5\%)} & {38 (60$\pm$10\%)} \\
	~~~cLIERs & {35 ($70^*\pm$12\%)} & {13 (81$^*\pm23$\%)} & {19 (50$^*\pm$12\%)} \\
	~~~eLIERs & {15 ($30^*\pm$8\%)} & {3 ($19^*\pm$11\%)} & {19 (50$^*\pm$12\%)} \\
	Seyfert & {1 (2$\pm$2\%)} & {3 (4$\pm$2\%)} & {3 (5$\pm$3\%)} \\
	\hline 
	Classifiable & {65 ($78^\dagger\pm10\%$)} & {83 ($100^\dagger\pm11\%$)} & {63 ($76^\dagger\pm10\%$)}\\
	Can't tell & {18 ($22^\dagger\pm5\%$)} & {0 ($0^\dagger\pm0\%$)} & {20 ($24^\dagger\pm5\%$)} \\
	\hline
	\end{tabular}
    \end{table}
  
  AGN feedback has been of significant interest in the extragalactic community as a potential cause of quenching for some years \citep[e.g.][]{Fabian2012,Heckman2014,Xu_2022}. While there is no clear result on links to AGN from the BPT analysis, another approach to determining whether AGN feedback may be directly influencing gas disc stability (i.e. preventing the cold gas in our main sample from forming stars) is via direct detection of winds or jets from galactic nuclei. Radio-mode feedback provides one avenue for stabilizing gas disks against collapse and star formation.  We cross match the three samples with the Faint Images of the Radio Sky at Twenty-cm (FIRST) survey \citep{Becker1995} to search for radio loud AGN. Using a sky position match within 10 arcseconds we find just {2} of the main sample are detected in FIRST, while {10} of the {Low HI} Control are detected and {16} of the high SF HI rich control. This analysis does not provide support for AGN feedback being the reason our HI rich main sample has usually low SFR. 
  
 Radio-quiet AGN may be another avenue for suppressing collapse in gas disks, but detecting these AGN would require a stacking analysis \citep[e.g.,][]{Roy2018} that we do not perform here. However, another direct AGN signature that has been studied for MaNGA galaxies are ``red-geysers" \citep{Cheung2016}, a subset of galaxies with evidence of bi-conical outflows in spatially resolved H$\alpha$ equivalent width maps. Further analysis of a larger sample by \citet{Roy2018} confirmed enhanced nuclear radio emission from red geysers consistent with weak AGN activity, suggesting red geysers are another manifestation of radio-quiet AGN feedback. \citet{Frank2022} explored whether red geysers are preferentially more gas-rich than other quiescent galaxies and found tentative evidence for a slight gas fraction enhancement, but not at a statistically significant level, making it unclear whether outflows are truly stabilizing gas effectively.
  
  Using existing red geyser classifications, we explore whether they are preferentially found in our gas-rich, non star-forming main sample. Red geysers, by their definition (being in red/low SF galaxies) are only identifiable in our main sample, and {Low HI} Control (as the {High SF} Control is composed of starforming galaxies which are almost all blue). We restrict this check to red galaxies with Near-Ultra-Violet (NUV) colors, NUV-$r>5$ and MPL-9 where red geysers are identified, which excludes the entire HI rich control sample. We find {3/25} and {7/42} of the galaxies in our Main and Low HI Control samples are red geysers. The 95\% binomial confidence intervals on the fraction of red geysers are {0.04-0.30} and  {0.08-0.31} for the Main and Low HI Control samples. Therefore, while there is a suggestion that we find fewer red geysers in the main sample compared the Low HI control, this is not statistically significant, and is definitely not providing evidence of an excess of red geysers among HI-rich low SF galaxies contributing to the suppression of SF, but rather perhaps a role in red geysers removing any HI gas which is present.

\subsection{Rotational Motion and Velocity Dispersion}
\begin{figure*}
    	\includegraphics[width=\columnwidth]{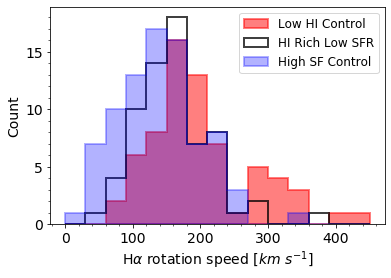}
    	\includegraphics[width=\columnwidth]{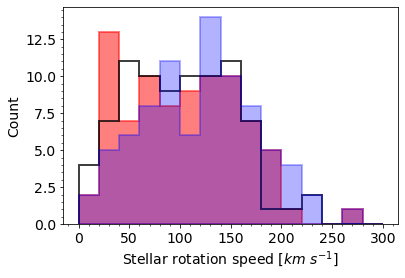}
    	\includegraphics[width=\columnwidth]{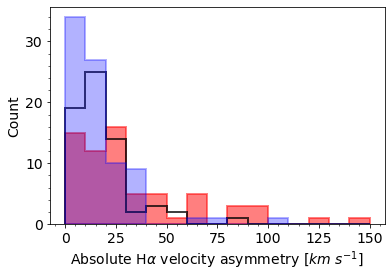}
    	\includegraphics[width=\columnwidth]{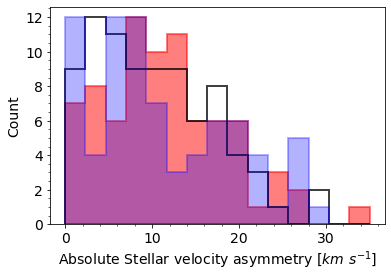}
    	\includegraphics[width=\columnwidth]{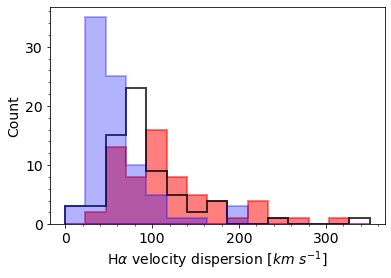}
    	\includegraphics[width=\columnwidth]{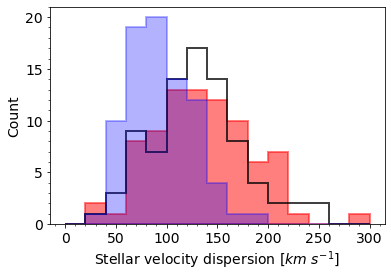}
        \caption{Distribution of properties measured from the H$\alpha$ velocity (left column) and stellar velocity (right column) fields. From top to bottom we show the bulk rotation, absolute asymmetry and velocity dispersion.  The main sample skews to larger values in all H$\alpha$ kinematics measures relative to the {Low HI} Control and also skews higher in the H$\alpha$ and stellar velocity dispersions, relative to the High SF Control (see Table \ref{tab:ks_table} for p-values).}
        \label{fig:velocity}
    \end{figure*}

 It is interesting to consider if the stars and gas in our HI rich low SF galaxies have similar motions to the control samples. We make histograms of both the \ha ~and stellar velocity dispersion, rotation widths and asymmetry (defined as the difference between {\tt HI\_CLIP} and {\tt LO\_CLIP} on both sides); all are taken directly from the DAP and measured within 1$r_e$. The histograms are presented in Fig.~\ref{fig:velocity}, while p-values from the Kolmogorov–Smirnov (KS) tests\footnote{KS tests performed using the SciPy Stats KS two-sample test package, \url{https://docs.scipy.org/doc/scipy/reference/generated/scipy.stats.ks_2samp.html}} comparing the main sample and the two controls are given in Table \ref{tab:ks_table}.
 
  \begin{table*}
	\centering
	\caption{KS p-values for comparison of all histograms in all Figures in this paper. Values which reveal significant differences ($p<0.05$) between the main sample and the controls are bolded. The Figure to which each row refers is given in the final column}
	\label{tab:ks_table}
	\begin{tabular}{lccr}
		\hline
		Parameter & {High SF Control} & {Low HI Control} & Figure\\
		\hline
		Stellar Mass & {0.998} & {0.586} & Fig.~\ref{fig:sample}\\
		HI Mass Fraction & {0.998} & {{\bf 0.000}} & Fig.~\ref{fig:sample}\\
		W2-W3 ($\sim$sSFR) & {{\bf 0.000}} & {0.998} & Fig.~\ref{fig:sample}\\
		\hline
		H$\alpha$ rotation speed & {0.170} & {{\bf 0.004}} & Fig.~\ref{fig:velocity} \\
		Absolute H$\alpha$ velocity asymmetry & {0.321} & {{\bf 0.002}}  & Fig.~\ref{fig:velocity} \\
		H$\alpha$ velocity dispersion (HA\_GSIGMA\_1RE) & {{\bf 0.000}} & {{\bf 0.000}} & Fig.~\ref{fig:velocity} \\
		
		Stellar rotation speed & {0.133} & {0.932} & Fig.~\ref{fig:velocity} \\
		Absolute Stellar velocity asymmetry & {0.091} & {0.353} & Fig.~\ref{fig:velocity} \\
		Stellar velocity dispersion (STELLAR\_SIGMA\_1RE) & {{\bf 0.000}} & {0.717} & Fig.~\ref{fig:velocity} \\
		\hline
		HI Line Width & {0.586} & - & Fig.~\ref{fig:HIwidths} \\
		\hline
		Disc galaxy or smooth (early-type)? \\
		({\tt t01\_smooth\_or\_features\_a02\_or\_disk\_debiased}) & {{\bf 0.018}} & {{\bf 0.001}} & Fig.~\ref{fig:morphology}\\
		Bulge size score & {{\bf 0.006}} & {0.053} & Fig.~\ref{fig:morphology}\\
		Visible spiral arms ({\tt t04\_spiral\_a08\_spiral\_debiased}) & {{\bf 0.015}} & {{\bf 0.001}} & Fig.~\ref{fig:morphology}\\
		Presence of a bar ({\tt t03\_bar\_a06\_bar\_debiased}) & {{\bf 0.032}} & {0.147} & Fig.~\ref{fig:morphology}\\
		\hline
		Metallicity (12 +$\log$[O/H]) & {{\bf 0.040}} & {0.189} & Fig.~\ref{fig:env-metal-tidal}\\
		Tidal Force from nearest neighbor, $Q$ & {0.961} & {{\bf 0.033}} & Fig.~\ref{fig:env-metal-tidal}\\
		Tidal Force from local environment, $Q_{\rm LSS}$ & {0.586} & {{\bf 0.006}} & Fig.~\ref{fig:env-metal-tidal}\\
        Local overdensity & {0.793} & {0.383} & Fig.~\ref{fig:env-metal-tidal}\\
		\hline
	\end{tabular}
    \end{table*}

 For \ha ~velocity measures, we only show galaxies with a median H$\alpha$ surface brightness within $1 r_e$ of more than $0.3 \times10^{-17}$ erg/s/cm$^2$/spaxel. This was an empircally determined limit which resulted in  reliable velocity measures from a well detected line across enough of the IFU that the velocity field was classifiable. This removes {17} galaxies from the main sample, and {15} from the Low HI Control (all of the H$\alpha$ maps from High SF Control meet this criteria). 

We find that there are statistically significant ($p<0.05$) differences between the main sample and the Low HI Control in all three \ha ~velocity measures - rotation speed, velocity asymmetry and velocity dispersion. In all three, the Low HI Control skews to larger values, possibly revealing a high outflow fraction caused by red geysers, or other non-rotational motion of the \ha ~gas (in the previous Section we demonstrated that the Low HI Control sample has a slight excess of red geysers, and a defining feature of red geysers is their high gas velocity dispersion; \citealt{Roy2018}).

While the \ha ~mean rotation and asymmetry are similar between the main sample and its High SF Control, we do find a significant difference in gas velocity dispersion, such that the High SF Control typically has lower gas dispersions, consistent with the \ha ~gas being more settled in this control sample. 

For the stellar velocity fields, we find that the only statistically significant difference is in stellar velocity dispersion between the main and High SF control samples, where we see a shift to larger velocity dispersions in the main sample relative to the High SF control. Again we find a picture where the High SF control show evidence for more settled disc like behaviour than the main sample. There is a hint that the {Low HI} control has even higher stellar velocity dispersion, but this is not a statistically significant difference ($p=0.717$). 
    
\begin{figure}
     	\includegraphics[width=\columnwidth]{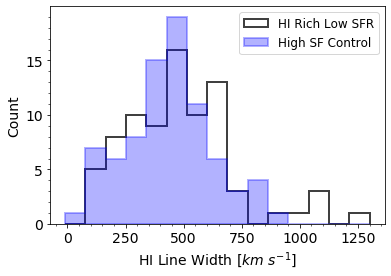}
          \caption{The distribution of inclination corrected HI line widths (in km/s) for galaxies in the main sample and high SF control (i.e. the two subsamples in which HI is detected). There is no statistically significant difference between these (see Table \ref{tab:ks_table} for p-values).}
         \label{fig:HIwidths}
 \end{figure}

     For the two HI detected samples we can also compare the HI rotation widths, corrected for inclination following the method described in \citet{Masters2019}. No notable difference is seen (see Figure \ref{fig:HIwidths}). 

\subsubsection{Kinematic Misalignment}\label{sec:kinematicalignment}
\begin{figure*}
     	\includegraphics[scale=0.6]{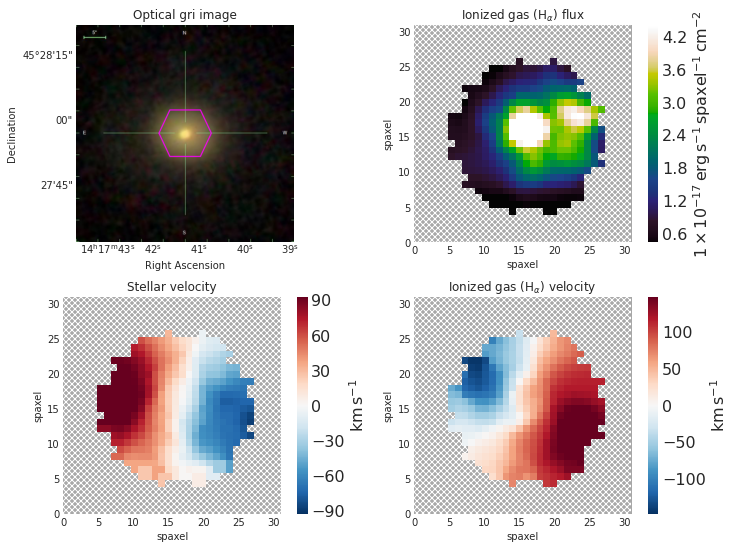}
          \caption{A set of example MaNGA maps showing a counter rotating galaxy (plateifu: 8329-1901) from HI rich low star-formation sample.  Panels show the optical gri image (upper left) with the purple hexagon indicating the region covered by the MaNGA bundle; the H$\alpha$ flux (upper right) in MaNGA data with the colour bar giving the scale and units; stellar and H$\alpha$ velocity maps (lower left and right respectively) with the colour bars showing the scale in km~s$^{_1}$ relative to the central velocity of the galaxy. The gas rotates in the opposite sense to the stars.}
         \label{fig:counterExample}
\end{figure*}

One of the cleanest tracers of external processes impacting galaxies is the signature of kinematic misalignment between stars and gas. Gas is likely to enter a galaxy from its local environment at a random angle relative to the rotation axis of any existing stars and gas, and therefore can set up kinematically distinct components. Visually inspecting the \ha ~and stellar velocity fields for all galaxies in the main and control samples we find a noticeable enhancement in the fraction of kinematic misalignment between gas and stars in our main sample of HI rich, but not star-forming galaxies. This is a subjective measure of kinematic asymmetry, but the enhancement is so large that a more quantitative method is unlikely to change our result. An example of a kinematically misaligned galaxy is shown in Figure \ref{fig:counterExample}. Overall we find that {10$\pm$3\%} of galaxies with both  stellar and gas velocity fields well measured (or {8}\% of all HI rich, low SF galaxies) have evidence of what are plausibly counter-rotating components, and another {10\%} show kinematic offsets. This adds up to {20\% (18\% of the sample)} showing clear evidence of kinematic misalignment between gas and stars. The frequency of kinematic offset appears similar in the {Low HI Control sample (10\%)}, but there are just {three} galaxies showing clear evidence of what might be counter-rotation in this sample. In the star-forming control sample we find no counter-rotators, and only {one} galaxies with evidence of any kind of kinematic offsets. This is a really notable difference between our sample and the controls and overall is a high fraction of kinematic misalignment. For a summary of these numbers see Table \ref{tab:kinematicMis_table}. One caveat to note is that we see a {similar} fraction of red geysers in our  {main sample of HI Rich Low SF galaxies ($4\pm2\%$) } and Low HI Control samples {($8\pm3\%$)}, so some of the gas velocities seen in the maps are plausibly outflows driven by the red geyser, rather than kinematic misalignment driven by gas inflow. {No red geysers are detected in our High SF Control (by definition red geysers are only possible to identify in low-SF samples).}

\begin{table}
	\centering
	\caption{Kinematic Misalignment fractions. All errors are {1-3\%} for these sample sizes. The range of fractions indicates comparison with the entire sample, or the subset with H$\alpha$ detections.}
	\label{tab:kinematicMis_table}
	\begin{tabular}{lccc} 
		\hline
		Parameter & {HI Rich Low SF}  & {High SF}  & {Low HI}  \\
		 & {Sample} & {Control} &  {Control} \\
		\hline
		H$\alpha$ Detected & {72 (87\%)} & {83 (100\%)} & {74 (90\%)} \\
		Counter-rotating &  {7 (8-10\%)} & {0 (0\%)} & {3 (4\%)} \\ 
		Offset Rotation & {8 (10-11\%)} & {1 (1\%)} & {8 (10-11\%)} \\ 
		\hline

	\end{tabular}
    \end{table}

In the \citet{Jin2016} analysis of the first year of MaNGA data, they find 5\% (66/1351) of the sample show kinematic misalignment. \citet{Xu2022} recently updated this work to make use of the almost complete MaNGA Sample, MPL-10, finding a similar overall fraction of 4.8\% or 456/9546 galaxies with kinematic misalignment. In \citet{Jin2016} they also show how the misaligned fractions vary as a function of stellar mass and SFR properties. They find the fraction of kinematic misalignment peaks (at around 10\%) at $\log M_\star/M_\odot = 10.5-11.0$.  The median stellar mass of our sample is $\log M_\star/M_\odot = {10.85}$ right in this peak (recall all our samples are arranged to have matching stellar mass distributions). They also find that misalignment is more common in more weakly SF galaxies. The upper sSFR limit for our low SF main and control sample is roughly $\log$(sSFR yr)  < -10.4, which is the lowest sSFR bin considered in \citet{Jin2016}, where they find misaligned fractions of 15-30\%. They do not investigate the cold gas properties of their sample, however it appears that our main sample has fractions of kinematic misalignment consistent with other low SF samples of this mass. Even with this, there is still perhaps evidence of a higher counter-rotating fraction than typical.   

\subsection{Morphology} \label{Sec:morphology}
The morphology of a galaxy is another way to trace its kinematics, and/or look for differences in the assembly history (bulge-dominated galaxies being more likely to have had significant mergers in the past). We made use of a catalog of morphologies of MaNGA galaxies which was released in the SDSS-IV DR16. These morphologies are mostly from Galaxy Zoo 2 \citep[GZ2,][]{Willet2013, Hart2016}, with some additional classification of MaNGA sample galaxies which were missing in GZ2. We identify spiral galaxies, using a selection of $p_{\rm features}>0.5$\footnote{where $p_{\rm features}$ is short for \\ {\tt t01\_smooth\_or\_features\_a02\_features\_or\_disk\_debiased}}. An additional cut of $b/a>0.3$ is used to exclude very edge on disc galaxies, so that we can make use of bar and spiral arm identifications in the remaining subset\footnote{Bar and spiral votes from {\tt t03\_bar\_a06\_bar\_debiased} and {\tt t04\_spiral\_a08\_spiral\_debiased} respectively}. We calculate the bulge size of these galaxies using Equation 4 of \citet{Masters2019}, which makes use of the consensus answers about bulge size in Galaxy Zoo.  

\begin{figure*}
    	\includegraphics[width=\columnwidth]{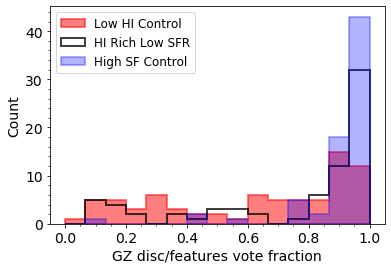}
    	\includegraphics[width=\columnwidth]{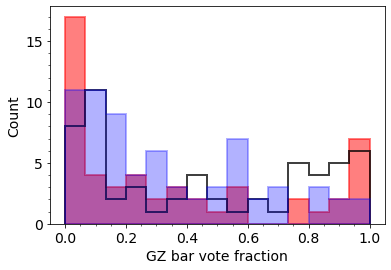}
    	\includegraphics[width=\columnwidth]{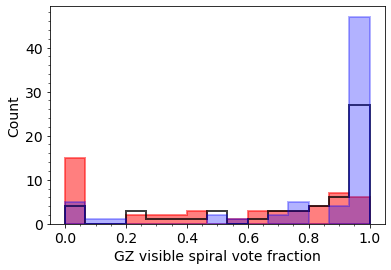}
    	\includegraphics[width=\columnwidth]{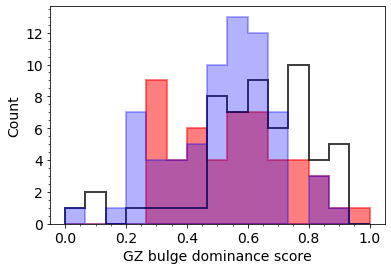}
        \caption{Histograms of morphological feature likelihoods from Galaxy Zoo debiased and weighted vote fractions for the three samples. Top Left: histogram of (debiased) votes for visible disc/features showing how the main sample has more discs that the {Low HI} Control, but fewer than the High SF Control. Top Right: for face-on disc galaxies, a histogram of (debiased) votes for a visible bar or (Lower Left) visible spirals. Lower right: a bulge size comparison (based on votes for bulge size - see text for details). The High SF Control is distinct in having both more spirals and smaller bulges than the main sample. The {Low HI} Control has notably fewer visible spirals. {Our HI Rich Low SF main sample also has significantly more visible bars than the High SF Control.} }
        \label{fig:morphology}
    \end{figure*}

Using Galaxy Zoo gives us continuous variables for all these morphologies, so we can make a comparison of the distribution of these morphological properties in the three samples (see Fig.~\ref{fig:morphology}, with p-values for KS tests comparing the distributions in Table \ref{tab:ks_table}). 

We find that our main sample contains galaxies which are morphologically different to those in both the high SF and low HI controls. The high SF control population is more likely to contain galaxies classified as a disc galaxy, with a smaller bulge, less visible bars, and more visible spiral arms than galaxies in the main sample. Galaxies in the low HI control are more likely to be classified as smooth, and if features are seen in galaxies in this control it is less likely that they are spiral arms than features seen in galaxies in the main sample.

Bars have {previously} been associated with low SF disc galaxies \citep{Masters2011} as well as disc galaxies with lower than average HI content \citep{Masters2012}. {We find a slight, but  statistically significant enhancement of bars in our sample over the High SF Control, while the Low HI Control and the main sample appear to have similar likelihood of hosting a visible bar. Overall 44$\pm$9\% of the main sample have a strong bar, ($p_{\rm bar}>0.5$ as defined in \citet{Masters2011}), while just 29$\pm$8\% of the High SF Control sample host strong bars.} 

Overall it appears that the presence (or absence) of spiral arms may be the most important feature. Our HI rich low SF main sample is found to be less likely to have very obvious arms {than} the High SF Control. Although it is true that both samples have the majority of galaxies in the $p_{\rm spiral} = 1$ bin, they are statistically different, as the main sample has more intermediate values of $p_{\rm spiral}$ and more galaxies in the $p_{\rm spiral} = 0$ bin than the high SF control. This is interesting as spiral arms have been invoked as a way to trigger SF (e.g. see \citet{SM22} for a recent review), although spirals may also be more visible in a disc with significant population of younger stars. Our main sample is also more likely to have visible spiral arms than the {Low HI} Control.

\subsection{Metallicity}

The gas phase metallicity provides a trace of how pristine the interstellar medium (ISM) is in these galaxies. We made use of MaNGA spectra along with Equation 5 of \citet{Kewly2002} to calculate the average gas metallicity of galaxies in our samples within 1 $r_e$ and consider if there are differences (see the Upper Left panel of Fig.~\ref{fig:env-metal-tidal} and Table \ref{tab:ks_table}). The distributions of gas phase metallicity are found to {be indistinguisbale between the main (HI Rich Low SF) and Low HI Control while differing only slightly} between the main and {the High SF Control sample. However, while statistically significant, the difference with the High SF Control is small and median values agree within the standard deviation}. This is not a clear signal of the physical origin of the HI Rich Low SF galaxies, but is suggestive of their formation from galaxies which could be in the Low HI Control with the addition of pristine gas, {which has not yet contributed to SF, so the gas phase metallicities have not yet changed.}

One caveat to note here is that we use the \citet{Kewly2002} prescription, which is based on an assumption of SF ionization to calculate metallicity for all spaxels, not just those with ionization from SF. As discussed in Section \ref{sec:ionization}, the different samples do have different amounts of non-SF based ionization. However, our use of [N II]/[O II] to estimate metallicity largely minimizes bias due to the fact that Nitrogen and Oxygen have very similar ionization energies, making their ratio highly insensitive to ionization parameter and the hardness of the ionizing spectrum \citep{Kewly2002, Zhang2017}. Therefore, we apply the same prescription on all spaxels regardless of whether they are primarily ionized by star formation.

\subsection{Environment}

Finally, since environment is known to be important in galaxy evolution \citep[e.g.][]{Peng2010}, we make use of a variety of environment measures to investigate both the local density and overdensities of our samples. 

\begin{figure*}
        \includegraphics[width=\columnwidth]{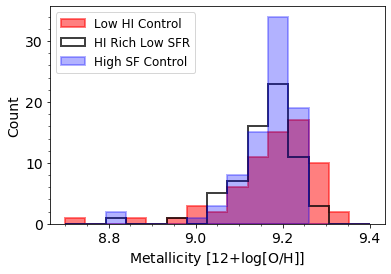}
        \includegraphics[width=\columnwidth]{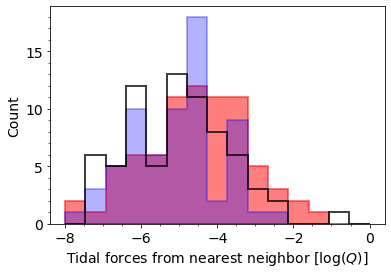}
    	\includegraphics[width=\columnwidth]{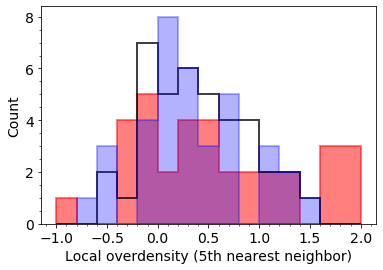}
        \includegraphics[width=\columnwidth]{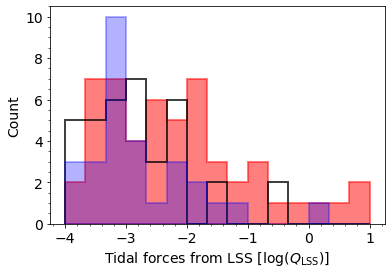}
        \caption{Histograms of the distribution of metallicity (Upper Left), {tidal force from the nearest neighbor, $Q$ (Upper Right on a log scale), local over density (to the 5th nearest neightbor, Lower Left), and tidal force from all nearby neighbors within 1 Mpc, Q$_{\rm LSS}$ (Lower Right) for all samples {See Section \ref{sec:other} for more on these)}. The tidal force parameters are unitless; larger values correspond to stronger tidal forces.} We find statistically different distribution of metallicity between the main and {High SF Control}. For the tidal force based environment measures ($Q$ and $Q_{\rm LSS}$) only the distributions between the {Low HI} Control and main sample are statistically different (see Table \ref{tab:ks_table} for p-values). }
        \label{fig:env-metal-tidal}
    \end{figure*}

We consider three measures of environment at different scales: a local overdensity (to the 5th nearest neighbor), tidal forces generated by large scale structures, $Q_{LSS}$; and tidal forces from the nearest neighboring galaxy, $Q$  (see Section \ref{sec:other} for more on all of these). Histograms of these values for all samples are shown in Fig.~\ref{fig:env-metal-tidal},  Table \ref{tab:ks_table} for KS p-values for the comparisons. We find a statistically significant difference in tidal forces (from both the LSS and the nearest neighbour) between the main sample and Low HI Control, such that the Low HI Control galaxies are more likely to be found in regions with higher tidal forces (larger values of $Q$ and $Q_{LSS}$). This control also appears to skew to higher local density values (although this difference is not found to be statistically significant). No difference is found in the environments of the main sample and High SF Control. 

\section{Discussion}\label{sec:discussion}
   
Our sample of unusual HI rich, yet low SF galaxies are significant outliers in the typical HI detected galaxy population, comprising just {$5\pm1$\%} of HI detected galaxies. Studying them in comparison to more typical galaxies (HI {Rich} and High SF, or {HI Poor and Low SF}) allows us to consider the evidence for or against likely physical origins of this subset of galaxies, which initially appeared to present interesting case studies for how quenching may occur in the galaxy population. 

Since we typically expect a strong correlation between HI content and star formation \citep[e.g.][]{Saintonge2016}, galaxies which are HI rich but lacking significant SF must therefore have experienced/be experiencing: 

\begin{enumerate}
    \item Recent accretion of HI gas (not yet impacting SF). 
    \item Heating of gas from feedback mechanisms blocking all or part of the HI$\rightarrow$H$_2\rightarrow$SF pipeline. 
    \item Some other process preventing or significantly slowing the cooling of the HI gas.
\end{enumerate}

In the case of recent accretion (i), we expect that the main sample would be closest in properties to the Low HI Control, just with the addition of HI. In addition we might expect to see (a) lower metallicities in the gas in the main sample relative to the Low HI Control; (b) an increase in the fraction of the main sample found in environments where there is more access to gas infall; greater disturbance in the H$\alpha$, or other gas velocity disturbances (e.g. counter-rotation). There is no statistically significant evidence that the gas phase metallicity of the main sample is lower than that of the Low HI Control, however the histogram of metallicity values in the low HI control visually looks skewed to higher metallicities (Figure \ref{fig:env-metal-tidal}). A future analysis with a larger sample size may reveal slight differences in the distribution (or confirm their similarly). There is a weak statistically significant correlation with environment measures -- the main sample is likely to be in lower density environments than the Low HI Control (Figure \ref{fig:env-metal-tidal}), where they may have better access to material for cold flow accretion.

 Perhaps most convincing as evidence of possible gas infall in the sample, there is a significant increase in the observed fraction of counter-rotation and kinematic offsets in our sample, suggesting that a recent gas accretion origin is plausible. For example, \citet{Starkenburg2019} use Illustris to consider the origin of galaxies with counter-rotation between stars and gas and suggest it requires gas infall onto a gas poor galaxy - explaining why galaxies with earlier type morphologies and lower SF tend to more often show counter-rotation. Both \citet{Bryant2019} and \citet{Ristea2022} find evidence which supports this model from samples of kinematically misaligned galaxies in the SAMI (Sydney-Australian-Astronomical-Observatory Multi-object Integral-Field Spectrograph) galaxy survey. \citet{Rathore2022} and \citet{Zhou2023} use MaNGA data to point to evidence that this kind of misaligned gas accretion is a formation mechanism for S0 galaxies and/or causing rejuvation of SF in previously quiescent S0s. These idea are consistent with us catching a sample at the early stages of cold gas infall onto a previously gas poor population. However there is no observable difference in the dynamical properties of the HI (see Figure \ref{fig:HIwidths}) which therefore appears to have had time to settle normally. There are a lot of open questions about what sets the rate of cool gas accretion onto galaxies, and factors which may influence this, including halo mass \citep[e.g.][]{White1978}, AGN activity \citep{Zinger2020}, large scale structure environment \citep{Liao2019}, even potentially the cosmic ray environment of the galaxies \citet{Butsky2020}. 

If the HI in our main sample does not represent recent gas infall, but rather some process blocking the gas-to-stars pipeline, either via heating (ii) or prevention of collapse/cooling (iii), there are multiple ways this can proceed including both internal and/or external processes. For example, cold gas can be heated, or prevented from cooling further by:  
\begin{itemize}
    \item AGN feedback (perhaps revealed via cLIERS)
    \item Supernova (SN) feedback, or other ionization from old stellar populations (which may be revealed in eLIERS)
    \item Dynamical heating via shocks (e.g. from bars, tidal interactions, counter-rotation). 
    \item Larger than typical dark matter fractions (i.e. halo heating)
    \item Being distributed in a very low density/high angular momentum gas disc. 
\end{itemize}

The latter model (of a high angular momentum, low density disc) has previously been used to explain large HI content in quiescent discs. For example, \citet{Zhang2019} found surprisingly large HI content in their sample of massive ($M_\star>10^{9.5}M_\odot$) quiescent disc galaxies. They suggest that it is remnant outer, high angular momentum HI which has a long timescale to migrate inwards in the absence of perturbations (although as we previously noted, \citet{Cortese2020} suggest that this may be due to them under-estimating the amount of extended SF in massive HI discs). \citet{Lemonias2014} obtained resolved HI maps of 20 massive, quiescent galaxies with large HI masses, finding the HI distributed at unusually large radii, suggesting low specific star formation rates may be caused by the low HI gas surface densities in the large HI discs. The HI-MaNGA data is unresolved so we have no way of obtaining the radial distribution of the HI. The sample investigated here may therefore be an interesting one to followup with resolved HI imaging in the future. Another interesting followup project would be looking at the UV extent of our HI Rich Low SF galaxies to search for low level extended SF in the low density disc. Extended UV discs have been observed to correlate well with large HI discs, and have previously been found in nearby galaxies \citep{Thilker2010, Cortese2012}.

If the main sample galaxies are experiencing feedback heating the gas, or preventing it from cooling, we might expect the process to reveal itself in how that sample differs from the normal high HI, high SF control. 

Considering the evidence for AGN feedback in our sample, we do see that about {4\%} of our main sample (and {8\%} of Low HI Control) host red geysers, and the cLIER fraction in the main sample is higher than in the High SF Control (see Section \ref{sec:ionization}). The large rates of kinematic misalignment (Section \ref{sec:kinematicalignment}) also could point in this direction, revealing the presence of gas outflows. Linking the two, \citet{Duckworth2020} presented mock MaNGA data based on galaxies in Illustris, and showed that those with kinematic misalignment also typically had enhanced BH growth and AGN activity. 

When considering SN feedback, or ionization from old stars, the High SF Control is found to host {smaller fraction of} eLIERs (Section \ref{sec:ionization}), {compared to the main sample}. So this process seems plausible. However we have not quantified the impact of the ionization from SF hiding the signature of eLIERS in the High SF Control, and the {Low HI} Control has similar fractions of both kinds of LIERs to the main sample. 

Dynamical heating of some kind seems to fit the evidence reasonably well, particularly given the high fractions of kinematic disturbance seen in the main sample compared to the High SF Control (Section \ref{sec:kinematicalignment}). In contrast, we see little difference in the environments of those two samples, no evidence for larger dynamical masses or halo masses (given the match on stellar mass and rotation widths), {but there is} evidence for a higher bar fraction (Section \ref{Sec:morphology}) in the main sample, which could cause shocks via radial flows. 

The main sample has less visible spiral arms than the High SF Control (Section \ref{Sec:morphology}), but this could be because spiral arms are enhanced by the presence of the SF more than that the lack of them inhibits SF. {It also has more visible bars, particularly strong bars, than the High SF Control, consistent with prior work which has associated low SF, in HI rich galaxies, with the presence of bars \citep{Masters2011,Masters2012}, however not all of the main sample has a bar, so this could only explain part of the sample.}

 Overall it seems that the sample is consistent with perhaps having some higher than typical recent cold gas infall, but could also be consistent with the presence of HI at large radii where it has been unable to cool. There is some evidence for an excess of AGN and/or other gas ionization which might additionally be blocking the gas-stars pipeline. 
 There is no clear single explanation why all galaxies in the sample selected are HI rich but have low SF, which supports a picture in which all of the possible processes could be happening in some part of the sample. Curiously \citet{Hallenbeck2014,Hallenbeck2016}, who studied very HI rich galaxies with typical SFR for their (high) stellar masses in the HIghMass sample also concluded that different formation processes were needed to explain these galaxies -- citing either recent accretion, or unusually high angular momentum in the HI disc. 

\section{Conclusions}

We study an unusual subset of MaNGA sample galaxies which are found to be HI detections while having very low sSFR. These galaxies represent just {5}\% of the HI detections of MaNGA galaxies in the large HI follow-up program for MaNGA \citep{Masters2019,Stark}. We construct control samples consisting of either both {low HI} low SFR or HI rich high SF galaxies, selected to have matching stellar mass distributions to the main sample in order to investigate the physical processes preventing SF from occurring. 

 As was previously found by \citet{Parkash2019}, HI rich galaxies with low star formation were found to have a significant proportion of LIERs in the MaNGA data. However unlike \citet{Parkash2019}, we find that our control sample of {low HI} with low SFR has a similarly high fraction of LIERs. This suggests that LIERs are specific to galaxies with low SFR, not necessarily high HI content. 
 
The availability of MaNGA data enables an investigation of spatially resolved BPT diagrams. Following \citet{Belfiore2016} LIERs were further categorized into extended-LIER (eLIER) and central-LIER (cLIER). The main, {HI Rich Low SF} sample was found to have a {moderately high} proportion of cLIERs {($70\%$) more comparable to High SF Control ($81\%$)} than Low HI Control {($50\%$)}. {Based on these results, we conclude that cLIER emission also favours high HI content (regardless of global SF).}
    
 MaNGA also provides velocity maps, both for stars and gas. We are only able to use \ha ~velocity maps for well detected \ha ~emission, but we find that the main, HI {Rich} Low SF sample has a higher H$\alpha$ velocity width, asymmetry and dispersion than the {Low HI} Control, as well as higher velocity dispersion when compared to the high SF control. The stellar velocity maps are much more similar between the three subsamples, with a difference only seen in the stellar velocity dispersion between the main and High SF Control.  
 
 One of the most notable differences between the main sample and the two controls is found in evidence of kinematic misalignment. We find that {about a fifth} of the galaxies in the main sample with both stellar and gas velocity maps show evidence for either counter-rotation, or kinematic misalignment. The {Low HI} Control shows a similar {but smaller} fraction of misalignment, but fewer counter-rotating galaxies, while the High SF Control shows much more regular velocity. 
 
  We also consider morphology, finding that the main sample is intermediate between the High SF Control (more discs/spirals) and the {Low HI} Control (more smooth galaxies). The High SF Control has much more visible spiral arms, {and is less likely to host a strong bar} than the main sample. 
  
 We look at metallicity, seeing {no clear evidence of any significant difference with either control. This suggests that if the main sample form from ‘low HI control-like’ galaxies by infall of pristine gas, this gas has not yet participated in enough SF to enrich the ISM.
 
 Finally we look at the environments of the three samples, using three different measures of environment on scales from the very local (nearest neighbour) to more large scale structure. No differences are found in the distribution of environments for the main and high SF control, while we find that the {Low HI} Control is slightly more likely to be found in higher density regions, perhaps more separated from intragalactic material available for cold flow accretion than the main sample. 
 
  We consider how these observations comparing the main sample with the two controls support various physical processes which could explain these unusual galaxies (Section \ref{sec:discussion}). No single physical explanation can explain all observations. We conclude that in some cases the HI may be distributed at large radii and therefore unable to cool to form $H_2$ which in turn would form new stars. Resolved HI imaging of a subset of this sample would therefore be interesting. Looking at the UV extent of this sample to seek low level extended SF in what might be a low density disc would also be interesting. In some cases we may be seeing recent gas infall. The high fraction of kinematic misalignment and counter-rotation is the best evidence for this model. We can also interpret some of the kinematic misalignment as outflows/heating driven by AGN or stellar feedback, which may also be playing a role in suppressing SF in these HI rich galaxies. 
  
  Galaxies are complex objects, and many physical processes are needed to explain their diversity of morphologies, masses, and star formation rates (e.g. see \citealt{KormendyKennicutt2004} for a review). This rare sample of HI rich, but low SF galaxies with resolved spectroscopy from the MaNGA survey presented an interesting puzzle alongside an opportunity to probe the processes quenching star formation in some galaxies.

\section*{Acknowledgements}

     This publication acknowledges Haverford College Koshland Integrated Natural Sciences Center (KINSC)'s support in summer 2020 and summer 2021 research program through a Summer Research Grant.\\ \\

     This publication makes use of the Sloan Digital Sky Survey (SDSS) database. Funding for the Sloan Digital Sky Survey IV has been provided by the Alfred P. Sloan Foundation, the U.S. Department of Energy Office of Science, and the Participating Institutions. 
     SDSS-IV acknowledges support and resources from the Center for High Performance Computing  at the University of Utah. The SDSS website is www.sdss.org. \\
     SDSS-IV is managed by the Astrophysical Research Consortium for the Participating Institutions of the SDSS Collaboration including the Brazilian Participation Group, the Carnegie Institution for Science, Carnegie Mellon University, Center for Astrophysics | Harvard \& Smithsonian, the Chilean Participation Group, the French Participation Group, Instituto de Astrof\'isica de Canarias, The Johns Hopkins University, Kavli Institute for the Physics and Mathematics of the Universe (IPMU) / University of Tokyo, the Korean Participation Group, Lawrence Berkeley National Laboratory, Leibniz Institut f\"ur Astrophysik Potsdam (AIP),  Max-Planck-Institut f\"ur Astronomie (MPIA Heidelberg), Max-Planck-Institut f\"ur Astrophysik (MPA Garching), Max-Planck-Institut f\"ur Extraterrestrische Physik (MPE), National Astronomical Observatories of China, New Mexico State University, New York University, University of Notre Dame, Observat\'ario Nacional / MCTI, The Ohio State University, Pennsylvania State University, Shanghai Astronomical Observatory, United Kingdom Participation Group, Universidad Nacional Aut\'onoma de M\'exico, University of Arizona, University of Colorado Boulder, University of Oxford, University of Portsmouth, University of Utah, University of Virginia, University of Washington, University of Wisconsin, Vanderbilt University, and Yale University.\\ \\

     This publication uses data products from the Wide-field Infrared Survey Explorer (WISE), which is a joint project of the University of California, Los Angeles, and the Jet Propulsion Laboratory/California Institute of Technology, funded by the National Aeronautics and Space Administration.\\ \\

     The paper uses data from observations \textcolor{black}{under project codes GBT16A\_095, GBT17A\_012, GBT19A\_127, GBT20B\_033 and GBT21B\_130} using the Green Bank Telescope (GBT). The Green Bank Observatory is a facility of the National Science Foundation operated under cooperative agreement by Associated Universities, Inc.\\ \\

     This publication uses the Marvin tool \citep{Cherinka2019} to generate maps of galaxy properties from MaNGA.

    This publication uses Galaxy Zoo 2 data generated via the Zooniverse.org platform, development of which is funded by generous support, including a Global Impact Award from Google, and by a grant from the Alfred P. Sloan Foundation.

%%%%%%%%%%%%%%%%%%%%%%%%%%%%%%%%%%%%%%%%%%%%%%%%%%
\section*{Data Availability}

All MaNGA data (including all the Value Added Catalogues used in this work) are publicly available via the SDSS website. Other public datasets are used, with links given in text.



%%%%%%%%%%%%%%%%%%%%%%%%%%%%%%%%%%%%%%%%%%%%%%%%%%

%%%%%%%%%%%%%%%%% APPENDICES %%%%%%%%%%%%%%%%%%%%%

%\appendix


% Don't change these lines
\bsp	% typesetting comment
\label{lastpage}
\end{document}